\documentclass[preprint,showpacs,preprintnumbers,amsmath,amssymb]{revtex4}
\usepackage{graphicx,epsfig,dcolumn,bm,epic,eepic,float}
\usepackage{amsmath}
\usepackage{latexsym}
\usepackage{color}
\usepackage{makeidx,shortvrb,latexsym}
\begin{document}
\unitlength 1 cm
\newcommand{\be}{\begin{equation}}
\newcommand{\ee}{\end{equation}}
\newcommand{\bearr}{\begin{eqnarray}}
\newcommand{\eearr}{\end{eqnarray}}
\newcommand{\nn}{\nonumber}
\newcommand{\vk}{\vec k}
\newcommand{\vp}{\vec p}
\newcommand{\vq}{\vec q}
\newcommand{\vkp}{\vec {k'}}
\newcommand{\vpp}{\vec {p'}}
\newcommand{\vqp}{\vec {q'}}
\newcommand{\bk}{{\bf k}}
\newcommand{\bp}{{\bf p}}
\newcommand{\bq}{{\bf q}}
\newcommand{\br}{{\bf r}}
\newcommand{\bR}{{\bf R}}
\newcommand{\up}{\uparrow}
\newcommand{\down}{\downarrow}
\newcommand{\fns}{\footnotesize}
\newcommand{\ns}{\normalsize}
\newcommand{\cdag}{c^{\dagger}}

\title {A new phenomenological Investigation  of $KMR$ and $MRW$ $unintegrated$ parton distribution functions}
\author{$M.$ $Modarres$$^\dag$}\altaffiliation {Corresponding author, Email :
mmodares@ut.ac.ir, Tel:+98-21-61118645, Fax:+98-21-88004781}
\author{$H. Hosseinkhani$$^\ddag$}
\author{$N. Olanj$$^\flat$}
\author{$M.R. Masouminia$$^{\dag}$}
 \affiliation{$^\dag$Physics Department, University
of Tehran, 1439955961, Tehran, Iran. } \affiliation{$^\ddag$Plasma
 and Fusion Research School, Nuclear Science and Technology
Research Institute, 14395-836, Tehran, Iran.\\}
\affiliation{$^\flat$Physics Department, Faculty of Science, Bu-Ali
Sina University, 65178, Hamedan, Iran.}
\begin{abstract}
The  longitudinal proton structure function, $F_L(x,Q^2)$,  from the
$k_t$ factorization formalism by using the unintegrated parton
distribution functions (UPDF) which are generated  through the
 KMR and  MRW procedures.
The  LO UPDF of the KMR prescription is extracted, by taking into
account the PDF of Martin et al, i.e. MSTW2008-LO and
 MRST99-NLO and next, the  NLO UPDF of
the MRW scheme is generated through the set of MSTW2008-NLO PDF as
the inputs. The different aspects of
 $F_L(x,Q^2)$ in the two approaches, as well as its perturbative
and non-perturbative parts are calculated. Then the comparison of
$F_L(x,Q^2)$ is made with the data given by the ZEUS and H1
collaborations. It is demonstrated that the extracted $F_L(x,Q^2)$
based on the UPDF of two schemes, are consistent to the experimental
data, and by a good approximation, they are independent to the input
PDF. But the one developed from the KMR prescription, have better
agreement to the data with respect to that of MRW. As it has been
suggested, by lowering the factorization scale or the Bjorken
variable in the related experiments, it may be possible to analyze
the present theoretical approaches more accurately.
\end{abstract}
\pacs{12.38.Bx, 13.85.Qk, 13.60.-r
\\ Keywords: $unitegrated$ parton
distribution, $DGLAP$ equation, splitting function} \maketitle
\section{Introduction}
In recent years, the extraction of $unintegrated$ parton
distribution functions ($UPDF$) have become very important, since
there exists plenty of
 experimental data on the various events, such as the
exclusive and  semi-inclusive
 processes in the high energy collisions in $LHC$, which indicates the
 necessity for computation of these $k_t$-dependent parton  distribution function.

The  $UPDF$, $f_a(x,k_t^2,\mu^{2})$, are the two-scale dependent
 functions, i.e.
 $k_t^2$ and $\mu^2$, which satisfy the $Ciafaloni$-$Catani$-$Fiorani$-$Marchesini$ ($CCFM$)
 equations \cite{4a,4b,4c,4d,4e}, where $x$, $k_t$ and
$\mu$ are  the longitudinal momentum fraction (the $Bjorken$
variable), the transverse momentum and the factorization scale,
respectively. They are
 $unintegrated$ over $k_t$ with respect to the conventional parton distributions ($PDF$) which satisfy
the  $Dokshitzer$-$Gribov$-$Lipatov$-$Altarelli$-$Parisi$ ($DGLAP$)
evolution equations  \cite{1a,1b,1c,1d}.

But the generation of $UPDF$ from the $CCFM$ equations is a
complicated task. So, in general,
  the Monte Carlo event generators \cite{6a,6b,6c,6d,6e,6f,6g,6h} are the main users of these equations.
  Since there is not a complete quark version of the $CCFM$ formalism, the alternative prescriptions
  are used for producing the quarks and  the gluons $UPDF$. Therefore, to obtain the $UPDF$,  $Kimber$, $Martin$ and $Ryskin$
  ($KMR$) \cite{8,tkimber} proposed a different procedure based on the standard $DGLAP$ equations in the leading order
  ($LO$)
  approximation, along with a modification due to the angular ordering condition, which is the key dynamical
  property of the $CCFM$ formalism. Later on,  $Martin$, $Ryskin$ and $Watt$ ($MRW$) extended
  the $KMR$ approach  for  the next-to-leading order ($NLO$) approximation \cite{watt,watt3,watt1},
  with this aim to improve the exclusive calculations.
  These two procedures  are the modifications to the standard $DGLAP$ evolution equations and can
 produce the $UPDF$ by using the $PDF$ as the inputs.

The general behavior and stability of the $KMR$ and $MRW$
prescriptions  were investigated in the references
\cite{mho,mh1,mh2,hm1,hm2}. Furthermore,   to check the reliability
of  generated $UPDF$, their relative behaviors were compared and
used to calculate the observable, deep inelastic scattering proton
structure function $F_2(x,Q^2)$. Then the predictions of these two
methods for the structure functions, $F_2(x,Q^2)$, were also
compared to the electron-proton deep inelastic measurements of $NMC$
\cite{thesis}, $ZEUS$ \cite{thesis1} and $H1+ZEUS$ \cite{hera}
experimental data.  The results were promising \cite{mho1}. It is
also concluded that \cite{mho1}, while the $MRW$ formalism is in
more compliance with the $DGLAP$ evolution equations  requisites,
but  it seems in the $KMR$ case, the angular ordering constraint
spreads the $UPDF$ to whole transverse momentum region, and makes
the results to sum up the leading $DGLAP$ and
$Balitski$-$Fadin$-$Kuraev$-$Lipatov$ ($BFKL$) logarithms
\cite{1n,2n,3n,4n,5n}.

Another important observable quantity in this connection is the
longitudinal structure function, i.e. $F_L(x,\mu^2)$, which is
proportional to the cross section of the longitudinal polarized
virtual photon with proton.  Particulary at small x, it is directly
sensitive to the gluon distributions i.e. $g\rightarrow q\bar{q}$
process. Moreover its  calculations in this region need the $k_t$
factorization formalism
\cite{7tkimber,7tkimber1,7tkimber2,7tkimber3,4kimberp}, which is
beyond the standard collinear factorization procedure \cite{proton}.
Recently, $Golec-Biernat$ and $Sta\acute{s}to$
\cite{stasto2,stasto3} ($GS$) have used the $k_t$ and collinear
factorizations
\cite{7tkimber,7tkimber1,7tkimber2,7tkimber3,4kimberp} as well as
the dipole approach to generate the longitudinal structure function,
but
 by using the $DGLAP$/$BFKL$ re-summation method,
developed by $Kwiecinski$, $Martin$ and $Stasto$ ($KMS$)
\cite{4kimber}, for  calculation of the $unintegrated$ gluon density
 at small x. They have parameterized the input  non-perturbative
 gluon distribution such that they could get the best fit to the
 experimental proton
 structure function data \cite{4kimber}.

 On the experimental side,  the longitudinal structure
 function has been measured by both the $H1$ \cite{H10,H101} and $ZEUZ$ \cite{ZEUS,ZEUS1}
 collaborations  at the $DESY$ electron-proton collider
 $HERA$. The $Q^2$ ranges have been varied between 12 to 90 and 24
 to 110 $GeV^2$ in each experiments, respectively.

As it was pointed out above, similar to our recent publication on
$F_2(x,Q^2)$ \cite{mho1}, in the present paper, we intend to
calculate $F_L(x,Q^2)$ by working in the the $k_t$-factorization
scheme. But rather than $KMS$ re-summation method pointed out above,
the $KMR$ and $MRW$ \cite{8,tkimber,watt,watt3,watt1} formalisms are
used to predict the $UPDF$ with the input $PDF$ of the
$MRST99$-$NLO$ \cite{MRST}, $MSTW2008$-$LO$ \cite{10} and
$MSTW2008$-$NLO$ \cite{10} which covers wide range of $(x,Q^2)$
plane. Then our results can be compared both with the experimental
data as well as the theoretical $KMS-GS$ presentation of
$F_L(x,Q^2)$. So the paper is organized as follows: In the section
$II$ we give a belief review of the $KMR$ and the $MRW$ formalisms
\cite{8,tkimber,watt,watt3,watt1} for extraction of the  $UPDF$ form
the phenomenological $PDF$ \cite{MRST,10}. The formulation of
$F_L(x,Q^2)$ based on the $k_t$-factorization scheme is given in the
section $III$. Finally, the section $IV$ is devoted to results,
discussions and conclusions.
\section{A brief review of the $KMR$ and the $MRW$ formalisms}
The $KMR$ and $MRW$ \cite{8,tkimber,watt,watt3,watt1,watt2} ideas
for generating the $UPDF$ work as follows: Using the given
integrated $PDF$ as the inputs, the $KMR$ and $MRW$ procedures
produce the $UPDF$ as their outputs. They are based on the $DGLAP$
equations along with some modifications due to the separation of
virtual and real parts of the evolutions, and the choice of the
splitting functions at leading order ($LO$) and the next-to-leading
order ($NLO$) levels, respectively:
\\ \\ ($i$) In the $KMR$ formalism \cite{8,tkimber}, the $UPDF$,
$f_{a}(x,k_{t}^2,\mu^{2})$ ($a=q$ and $g$), are defined in terms of
the quarks and the gluons $PDF$, i.e.:
\begin{eqnarray}
f_{q}(x,k_{t}^2,\mu^{2})=T_q(k_t,\mu)\frac{\alpha_s({k_t}^2)}{2\pi}
\int_x^{1-\Delta}dz\Bigg[P_{qq}(z)\frac{x}{z}\,q\left(\frac{x}{z} ,
{k_t}^2 \right)+ P_{qg}(z)\frac{x}{z}\,g\left(\frac{x}{z} , {k_t}^2
\right)\Bigg],
 \label{eq:8}
\end{eqnarray}
and
\begin{eqnarray}
f_{g}(x,k_{t}^2,\mu^{2})=T_g(k_t,\mu)\frac{\alpha_s({k_t}^2)}{2\pi}
 \int_x^{1-\Delta}dz\Bigg[\sum_q
P_{gq}(z)\frac{x}{z}\,q\left(\frac{x}{z} , {k_t}^2 \right) +
P_{gg}(z)\frac{x}{z}\,g\left(\frac{x}{z} , {k_t}^2 \right)\Bigg],
 \label{eq:9}
\end{eqnarray}
respectively, where, $P_{aa^{\prime}}(x)$, are the $LO$ splitting
functions,  and the survival probability factors, $T_a(k_t,\mu)$,
are evaluated from:
\begin{eqnarray}
T_a(k_t,\mu)=\exp\Bigg[-\int_{k_t^2}^{\mu^2}\frac{\alpha_s({k'_t}^2)}{2\pi}\frac{{dk'_t}^{2}}{{k'_t}^{2}}
 \sum_{a'}\int_0^{1-\Delta}dz'P_{a'a}(z')\Bigg].
 \label{eq:5}
\end{eqnarray}
The angular ordering condition ($AOC$) \cite{3a,3b}, which is a
consequence of coherent emission of gluons, on the last step of the
evolution process \cite{watt2}, is imposed. The $AOC$ determined the
cut off, $\Delta=1-z_{max}=\frac{k_t}{\mu+k_t}$,  to prevent $z=1$
singularities in the splitting functions, which arises from the soft
gluon emission. As it has been pointed out in the references
\cite{8,tkimber}, the $KMR$ approach has several main
characteristics. The important one, is the existence of the cut off
at the upper limit of the integrals, that makes the distributions to
spread smoothly to the region in which $k_t>\mu$ i.e. the
characteristic of the small $x$ physics, which is principally
governed by the $BFKL$ evolution \cite{1n,2n,3n,4n,5n}. This feature
of the $KMR$, leads to the $UPDF$ with the behavior very similar to
the unified $BFKL$+$DGLAP$ formalism \cite{8,tkimber}. The $UPDF$
based on the $KMR$ formalism, have been widely used in the
phenomenological
calculations which depend on the transverse momentum \cite{1,2,3,4,5,6,7,81,9,101,11,12}.  \\
($ii$) In the $MRW$ formalism \cite{watt,watt3,watt1}, the similar
separation of real and virtual contributions to the $DGLAP$
evolution is done, but the procedure is performed at the $NLO$ level
i.e.,
\begin{eqnarray}
f_{a}^{NLO}(x,k_{t}^2,\mu^{2})=\int_x^{1}dz
T_a(k^{2},\mu^{2})\frac{\alpha_s({k}^2)}{2\pi}
\sum_{b=q,g}P_{ab}^{(0+1)}(z)\,b^{NLO}\left(\frac{x}{z} , {k}^2
\right)\Theta(\mu^{2}-k^{2}),
 \label{eq:10}
\end{eqnarray}
where
\begin{eqnarray}
P_{ab}^{(0+1)}(z)=P_{ab}^{(0)}(z)+\frac{\alpha_s}{2\pi}P_{ab}^{(1)}(z),k^2=\frac{k_t^2}{1-z}
 \label{eq:11}.
\end{eqnarray}
In the  equations (\ref{eq:10}) and (\ref{eq:11}) the $P_{ab}^{(0)}$
and the $P_{ab}^{(1)}$ denote the $LO$ and the $NLO$ contributions
of the splitting functions, respectively. It is  obvious from
equation (\ref{eq:10}) that in the $MRW$ formalism, the $UPDF$ are
defined such that to ensure $k^{2}<\mu^2$. Also,  the survival
probability factor, $T_a(k^{2},\mu^{2})$,  are obtained as follows:
\begin{eqnarray}
T_a(k^{2},\mu^{2})=\exp\Bigg(-\int_{k^2}^{\mu^2}\frac{\alpha_s({\kappa}^2)}{2\pi}\frac{{d\kappa}^{2}}{{\kappa}^{2}}
 \sum_{b=q,g}\int_0^{1}d\zeta \zeta
P_{ba}^{(0+1)}(\zeta)\Bigg),
 \label{eq:12}
\end{eqnarray}
where $P_{ab}^{(i)}$ (which is singular in the $z\rightarrow1$) is
given in the reference \cite{19w}. $MRW$ have  demonstrated that the
sufficient accuracy can be obtained by keeping only the $LO$
splitting functions together with the $NLO$ integrated parton
densities. So, by considering angular ordering, we can use $P^{(0)}$
instead of $P^{(0+1)}$. As it is mentioned above unlike the $KMR$
formalism, where the angular ordering is imposed to the all of terms
of the equations (\ref{eq:8}) and (\ref{eq:9}), in the $MRW$
formalism, the angular ordering is imposed to the terms in which the
splitting functions are singular, i.e. the terms that include
$P_{qq}$ and $P_{gg}$.
\section{The formulation of $F_L(x,Q^2)$ in the $k_t$-factorization approach}
The $k_t$-factorization approach has been discussed in the several
works i.e. references \cite{7tkimber,4c,7tkimber3,new7,new8}. In the
following equation \cite{12kimber,13kimber,14kimber,stasto2}, the
different terms i.e the perturbative and the non-perturbative
contributions to the $F_L(x,Q^2)$ has been broken into the sum of
gluons from the quark-box (the first term i.e. the $k_t$
factorization part), see figure 1 \cite{watt1}), quarks (the second
term) and the non-perturbative gluon  (the third term) Parts:
$$
F_L (x,Q^2) =\Bigg[\frac{Q^4}{\pi^2}\sum_{q} e_q^2
 \int\frac{dk_t^2}{k_t^4} \Theta(k^2-k_0^2)\int_0^{1}d\beta\int
d^2\kappa_ t \alpha_s(\mu^2) \beta^2(1-\beta)^2\left
(\frac{1}{D_1}-\frac{1}{D_2}\right)^2\times$$
$$
f_g\left(\frac{x}{z},k_t^2,\mu^2\right)\Theta(1-{x\over z})\Bigg] +
 \Bigg[\frac{4}{3} \int_x^{1}\frac{dy}{y}
\frac{\alpha_s(Q^2)}{\pi} (\frac{x}{y})^2 F_2 (y,Q^2)\Bigg]+$$
\begin{eqnarray}
\frac{\alpha_s(Q^2)}{\pi} \Bigg[\sum_{q} e_q^2
\int_x^{1}\frac{dy}{y}(\frac{x}{y})^2 (1-\frac{x}{y}) y
g(y,k_0^2)\Bigg] \label{eq:2p},\ \ \ \ \ \ \ \ \ \ \ \ \ \ \ \ \ \ \
\ \ \ \ \
\end{eqnarray}
where the second term is (see \cite{stasto4,stasto5} ):
\begin{eqnarray}
\sum_{q} e_i^2 \frac{\alpha_s(Q^2)}{\pi} \frac{4}{3}
\int_x^{1}\frac{dy}{y}(\frac{x}{y})^2
[q_i(y,Q^2)+\overline{q}_i(y,Q^2)]\nonumber .
\end{eqnarray}
In the above equation, in which the graphical representations of
$k_t$ and $\kappa_t$ have been introduced in the figure 1,  the
variable $\beta$ is defined as the light-cone fraction of the photon
momentum carried by the internal quark \cite{new8}. Also, the
denominator factors are:
\begin{eqnarray}
D_1&=&\kappa_t^2+\beta(1-\beta)Q^2+m_q^2,\nonumber\\D_2&=&({\bf{\kappa_t}}
-{\bf{k_t}})^2 +\beta(1-\beta)Q^2+m_q^2
 \label{eq:3p}.
\end{eqnarray}
Then by defining ${\bf\kappa^\prime_t}={\bf\kappa_t}-(1-\beta){\bf
k}_t$, the variable y takes the following form:
$$
y=x(1+{{{\kappa^\prime}^2+m^2_q}\over{\beta(1-\beta)Q^2}}),
$$
and
\begin{eqnarray}
\frac{1}{z}=1+\frac{\kappa_{t}^{2}+m_q^2}{(1-\beta)Q^2}+\frac{k_t^2+\kappa_t^2-2{\bf{\kappa_t}}.{\bf{k_t}}+m_q^2}{\beta
Q^2}
 \label{eq:a}.
\end{eqnarray}
As in the reference \cite{4kimber}, the scale $\mu$ which controls
the $unintegrated$ gluon and the $QCD$ coupling constant
$\alpha_s$, is chosen as follows:
\begin{eqnarray}
\mu^2=k_t^2+\kappa_t^2+m_q^2
 \label{eq:b}.
\end{eqnarray}
One should note that the coefficients used for quark and
non-perturbative gluon contributions depend on the transverse
momentum.  As it has been briefly explained before, the main
prescription for $F_L$ consists of three terms; the first term is
the $k_t$ factorization which explains the contribution of the
$UPDF$ into the $F_L$. This term is derived with the use of pure
gluon contribution. However, it only counts the gluon contributions
coming from the perturbative region, i.e. for $k_t>1$ $GeV$ , and
does not have anything to do with the non-perturbative
contributions. In the reference \cite{stasto4}, it has been shown
that a proper non-perturbative term can be derived from the $k_t$
factorization term, compacting the $k_t$ dependence and the
integration with the use of a variable-change, i.e. y, that carries
the $k_t$ dependence. Nevertheless, there is a calculable quark
contribution in the longitudinal structure function of the proton,
which comes from the collinear factorization, i.e. the second term
of the equation (\ref{eq:2p}).

For the charm quark, $m$ is taken to be $m_c=1.4 GeV$, and $u$, $d$
and $s$ quarks masses are neglected. We also use the same
approximation to save the computation time \cite{tkimber}, the one
we did for the calculation of $F_2(x,Q^2)$ \cite{mho1} i.e the
representative "average" value for $\phi$, $\langle \phi \rangle
=\frac{\pi}{4}$ for perturbative gluon contribution. This
approximation has been checked in the reference \cite{tkimber} (page
83). The rest of $\phi$ angular integration can be performed
analytically by using a series of integral identities given in the
reference \cite{anal}. We will also verify this approximation in the
next section. The $unintegrated$ gluon distributions are not defined
for $k_t$ and $\kappa_t <k_0$, i.e. the non-$perturbative$ region.
So, according to the reference \cite{12kimber}, $k_0$ is chosen to
be about one $GeV$ which is around the charm mass in the present
calculation, as it should be. On the other hand, one expects that
the discrepancy between the $k_t$-factorization calculation and the
experimental data can be eliminated by using the $PDF$, which have
been fitted to the same data for $F_2(x,Q^2)$  \cite{stasto1} with
respect to the re-summation method of $KMS$ \cite{4kimber}.
\section{Results, discussions and conclusions}
In the figure 2, the longitudinal proton structure functions in the
frameworks of  $KMR$ (left panels) and $MRW$ (right panels)
formalisms, by using the MRST99 \cite{MRST}
 and  the $MSTW2008-NLO$  \cite{10}  $PDF$
 inputs, versus x,  for $Q^2$=$2, 4, 6, 12$ and $15$ $GeV^2$ are
 plotted, respectively. Their total $F_L(x,Q^2)$  and the contributions from $k_t$ factorization
scheme, the quarks and the no-perturbative parts (see the equation
(\ref{eq:2p}) are presented with different curve styles. The
behavior of  $F_L(x,Q^2)$ mostly comes from the $k_t$ factorization
contribution especially as the $Q^2$ is increased and it is more
sizable in case of $MRW$ approach. By rising up the $Q^2$ values the
contribution of the $k_t$-factorization becomes dominant. Another
point is the decrease of non-perturbative parts at small  x, in the
case of $MRW$ scheme. As we discussed in our pervious works, this is
expected. Since the $KMR$ constraint spreads the $UPDF$ to the whole
transverse momentum region \cite{mho1} and it sums up the both
leading $DGLAP$ and $BFKL$ logarithms contributions. The general
behavior of two schemes in the figure 2 shows some differences also
at lower $Q^2$ scales, while the values and behaviors of quarks and
$k_t$-factorization portions in both formalisms are almost similar,
the non-perturbative contributions have more different values and
behavior in the $x\simeq 0.01$. The later point plays the main role
in the discrepancies of the total $F_L(x,Q^2)$ at lower $Q^2$. On
the other hand the non-perturbative contribution in each case
remains almost fix through the variation of $Q^2$. These effects
have root in the parent PDF sets at non-perturbative boundary which
is very sensitive to the discipline and procedure of the $PDF$
generating group.This figure can also be compared with the figure 2
of $GS$ \cite{stasto2} at $Q^2$=$2, 4$ and $6$ $GeV^2$. There are
general agreements between our approaches and those of $GS$, which
have used the $DGLAP$/$BFKL$ re-summation method, developed by
$Kwiecinski$, $Martin$ and $Stasto$ ($KMS$) \cite{4kimber}, for
calculation of the $unintegrated$ gluon density
 at small x. This agreement is more visible at larger $Q^2$ and in
 the $KMR$ approach, which is expected. However our longitudinal proton structure
 function results go smoothly to zero with respect to those $GS$ as x becomes larger. The
 reason  comes from both our input $PDF$, which is valid for the
 whole $(x,Q^2)$ plane, and the calculation of $UPDF$ which are
 calculated by using the $KMR$ and $MRW$ approaches, which are
 full fill the $DGLAP$ requirements.

Our longitudinal proton structure function results for larger values
of $Q^2$, with the different input $PDF$ i.e. $MERST99$ \cite{MRST},
$MSTW2008-LO$ (using $KMR$ formalism)  and $MSTW2008-NLO$ \cite{10}
(using $MRW$ formalism)  are given the figures 3, 4 and 5,
respectively. Again the total $F_L(x,Q^2)$  and the contributions
from $k_t$ factorization scheme, the quarks and the no-perturbative
parts are presented with different curve styles. The results are
mostly decreasing function x, for the various values of $Q^2$. There
are sizable differences between the $MERST99$ and $MSTW2008-LO$. On
the other hand,  as one should expect, for large value of $Q^2$ the
results of the $KMR$ and $MRW$ behave more similarly. As we pointed
out before, again the $k_t$ factorization contributions are
dominant. The increase in the values of $F_L(x,Q^2)$ in the figure 4
is due to the increase of the input PDF at LO approximation. The
reason that the results of $F_L(x,Q^2)$ approach to same values as
$x$ and $Q^2$ increases, which is a heritage of the parent $DGLAP$
evolution.

In order to analyze the above $Q^2$ dependent more clearly, in the
figure 6, the longitudinal proton structure functions are plotted
against $Q^2$ for two different values of $x=0.001$ and $0.0001$.
Note, that for large $Q^2$, especially the $MRW$ approach, needs
large computation time. So we have stopped at $Q^2=100$ $GeV^2$ for
this procedure. There are sizable differences between the two
approaches and results coming from the two different input $PDF$ .
But this should not be very important regarding the experimental
data, that we will discuss later on. In the figure 7, a comparison
is made between the three different,  $F_L(x,Q^2) $ results, namely
$KMR$ procedure with $MERST99$ and $MSTW2008-LO$ inputs and $MRW$
scheme with $MSTW2008-NLO$ inputs. Especially there are large
differences between $KMR$ and $MRW$ approaches at large $Q^2$. The
above results can be directly compared that of $GS$ \cite{stasto2}
(see their figure 3). Very similar behavior is observed especially
between the $k_t$ factorization approaches.

In the figures 8,  9 and 10, we present our results in the range of
energy available in the $H1$ and $ZEUS$ data \cite{hera},
respectively. Note that for $Q^2\geq 80$ $GeV^2$,  because of large
computation time, we have only given  four points (filled squares)
for the $MRW$ case. Very good agreements is observed between our
result and those of experimental data at different $Q^2$ and x
values. It seems with present existed data the $UPDF$ of gluons
generated with different input $PDF$ and constraints procedures, one
can reasonably explain the $H1$ and the $ZEUS$ experimental data. It
looks that even at low energies and small x values (see the figure
8); we find good agreement between our calculation and available
data. However, as we mentioned before and it has been stated by
several authors, the $F_L$ is mainly driven through the gluons
distributions, especially at low values of x. The fact that $F_2$ is
not accurately fit the data (see our previous
 work \cite{mho1}), but we get good agreement between the $F_L$
calculations and H1 and ZEUS data, could be caused of the
quark-quark contributions which has more contribution to $F_2$.
Since $F_L$ is more sensitive to the gluons $UPDF$ with respect to
$F_2$. So one can conclude that present calculation can confirm that
the $KMR$ and $MRW$ procedures (for generating the gluon $UPDF$) and
the $k_t$-factorization scheme can reproduce reasonable $F_2$
(considering our previous work \cite{mho1})  and present $F_L$. On
the other hand, as we stated previously:(1)Present results also
shows good agreement with the theoretical calculations of $GS$,
which have used more complicated approach such as $KMS$. (2)It is
interesting that the $KMR$ and $MRW$ $UPDF$ can generate reasonable
$F_L$ without using any free parameter  in the (x, $Q^2$)-plane even
at low $Q^2$ (regarding figure 8), especially the $UPDF$ generated
for gluons.

Finally,  the verification of the fact  that the $\phi$ integration
of perturbative gluon contribution can be averaged  by setting
$<\phi>=\pi/4$, which was discussed in the end of previous section,
is presented in the figure 11, for four values of $Q^2= 3.5, 12, 60$
and $110$ $Gev^2$ by using the $KMR$ formalism and the $MRST99$. It
is clearly seen that the above approximation does work properly and
one can save much computation time.

In conclusion,   the longitudinal proton structure functions,
$F_L(x,Q^2)$, were calculated  based on the $k_t$ factorization
formalism, by using the $UPDF$ which are generated  through the
$KMR$ and
 $MRW$ procedures.  The $LO$ $UPDF$ of the $KMR$ prescription is extracted, by taking into
account the $PDF$ of $MSTW2008$-$LO$ and $MRST99$-$NLO$ and also,
the $NLO$ $UPDF$ of the $MRW$ scheme is generated through the set of
$MSTW2008$-$NLO$ $PDF$ as the inputs. The different aspects of the
$F_L(x,Q^2)$ in the two approaches, as well as its perturbative and
non-perturbative parts were calculated and discussed. It was shown
that our approaches are in agreement with those given $GS$. Then the
comparison of $F_L(x,Q^2)$ was made with the data given by the
$ZEUS$ and $H1$ collaborations at $HERA$. It was demonstrated that
the extracted longitudinal proton structure functions based on the
$UPDF$ of above two schemes, were consistent with the experimental
data, and by a good approximation, they are independent to the input
$PDF$. But as it was pointed out in our previous work \cite{mho1},
the one developed from the $KMR$ prescription, have better agreement
to the data with respect to that of $MRW$. Although the $MRW$
formalism is in more compliance with the $DGLAP$ evolution equations
requisites, but  it seems in the $KMR$ case, the angular ordering
constraint spreads the $UPDF$ to whole transverse momentum region,
and makes the results to sum up the leading $DGLAP$ and $BFKL$
logarithms. At first, it seems that there should be  a theoretical
support for applying the angular ordering condition  only to the
diagonal splitting functions, in accordance with reference
\cite{watt1}. But as it has been mentioned in the references
\cite{mho1,mmho1}, this  phenomenological modifications of the $KMR$
approach (including the application of the $AOC$  to all splitting
functions) works as an "effective model"  that spreads the $UPDF$ to
the $k_t > \mu$ (a characteristic of low x physics) which enables it
to represent a good level of agreement with the data.  Beside this
in our new work \cite{mmho1} in which we have  calculated the $F_L$
in the dipole approximation according to the LO prescription of
reference \cite{watt1},  it is  shown that there is not much
difference if one applies the $AOC$ to the all splitting functions
i.e. to use the $KMR$ $UPDF$ instead of using LO prescription of
reference \cite{watt1}. On the other hand, in this paper we have
focused on comparison of the LO and the $NLO$ calculation of $F_L$
and since the calculations are very time consuming we restricted the
results to the $LO-KMR$ and $NLO-MRW$.

As it has been suggested in the  reference \cite{stasto2}, by
lowering the factorization scale or the $Bjorken$ variable in the
experimental measurements, it may be possible to analyze the present
theoretical approaches more accurately.

\begin{acknowledgements}
$MM$ would like to acknowledge  the Research Council of University
of Tehran and Institute for Research and Planning in Higher
Education for the grants provided for him.
\end{acknowledgements}

\begin{figure}[ht]
\caption{The quarks-box and exchanged diagrams in the photon-gluon
fusion process discussed in the $k_t$ factorization formula in the
text.}
\end{figure}
\begin{figure}[ht]
\caption{The longitudinal proton structure functions in the
frameworks of  $KMR$ (left panels, using the MRST99 $PDF$ data as
inputs) and $MRW$ (right panels, using the $MSTW2008-NLO$ data as
inputs) $UPDF$, versus x,  for  $Q^2$=$2, 4, 6, 12$ and $15$
$GeV^2$. Their total value  and the contributions of $k_t$
factorization scheme, the quarks and the no-perturbative parts  are
presented with different curve styles.}
\end{figure}
\begin{figure}[ht]
\caption{The longitudinal proton structure functions in the
frameworks of  $KMR$ by using  the $MRST99$ $PDF$ data versus x, for
$Q^2$=$12, 15, 20, 25, 35, 45, 60, 80, 90$ and $110$ $GeV^2$. Their
total value  and the contributions of $k_t$ factorization scheme,
the quarks and the no-perturbative parts  are presented with
different curve styles. }
\end{figure}
\begin{figure}[ht]
\caption{The longitudinal proton structure functions in the
frameworks of  $KMR$ by using  the $MSTW2008-LO$ $PDF$ data versus
x, for $Q^2$=$12, 15, 20, 25, 35, 45, 60, 80, 90$ and $110$ $GeV^2$.
Their total value  and the contributions of $k_t$ factorization
scheme, the quarks and the no-perturbative parts  are presented with
different curve styles. }
\end{figure}
\begin{figure}[ht]
\caption{The longitudinal proton structure functions in the
frameworks of  $MRW$ and  by using  the  $MSTW2008-NLO$ $PDF$ data
versus x, for $Q^2$=$12, 15, 20, 25, 35$ and $45$ $GeV^2$. Their
total value and the contributions of $k_t$ factorization scheme, the
quarks and the no-perturbative parts  are presented with different
curve styles. }
\end{figure}
\begin{figure}[ht]
\caption{The longitudinal proton structure functions in the
frameworks of  $KMR$ and $MRW$ by using  the  $MRST99$,
$MSTW2008-LO$ and $MSTW2008-NLO$ $PDF$ data versus $Q^2$ ($GeV^2$),
for fix x=0.001 and 0.0001. Their total values and the contributions
of $k_t$ factorization scheme, the quarks and the no-perturbative
parts are presented with different curve styles. }
\end{figure}
\begin{figure}[ht]
\caption{The comparison of total  longitudinal proton structure
functions in the frameworks of  $KMR$ and $MRW$ by using  the
$MRST99$, $MSTW2008-LO$ and $MSTW2008-NLO$  $PDF$ data versus $Q^2$
($GeV^2$), for the fix x=0.001 and 0.0001.}
\end{figure}
\begin{figure}[ht]
\caption{The comparison of total  longitudinal proton structure
functions, in the frameworks of  $KMR$ and $MRW$ by using  the
$MRST99$, $MSTW2008-LO$ and $MSTW2008-NLO$ $PDF$ data versus x at
$Q^2$=$2, 2.5, 3.5, 5, 6.5, 8.5 $ and $9$ $GeV^2$, with the
corresponding $ZEUS$ and $H1$ data (filled-triangles and bold
points), respectively.}
\end{figure}
\begin{figure}[ht]
\caption{The comparison of total  longitudinal proton structure
functions, in the frameworks of  $KMR$ and $MRW$ by using  the
$MRST99$, $MSTW2008-LO$ and $MSTW2008-NLO$ $PDF$ data versus x at
$Q^2$=$12, 15, 20, 25, 35, 45, 60$ and $90$ $GeV^2$, with the
corresponding $H1$ data (bold points).}
\end{figure}
\begin{figure}[ht]
\caption{The comparison of total  longitudinal proton structure
functions, in the frameworks of  $KMR$ and $MRW$ by using  the
$MRST99$, $MSTW2008-LO$ and $MSTW2008-NLO$ $PDF$ data versus x at
$Q^2$=$24, 32, 45, 80$ and $110$ $GeV^2$, with the corresponding
$ZEUS$ and  $H1$ data (filled-triangle and bold points),
respectively.}
\end{figure}
\begin{figure}[ht]
\caption{The comparison of perturbative gluon contribution to $F_L$
by performing the $\phi$ integration (exact) and the approximated
one with $\phi=\pi/4$ , in the frameworks of $KMR$  by using the
$MRST99$  versus x at $Q^2$=$3.5, 12, 60$ and $110$ $GeV^2$.}
\end{figure}
\end{document}